%% file: proceedings.tex
\documentclass[12pt]{article}
\usepackage{epsfig}

\input econfmacros.tex
\def\Title#1{\begin{center} {\Large {\bf #1} } \end{center}}

\begin{document}

\Title{Kaons and Hyperons rare decays by the NA48 experiment at CERN}

\bigskip\bigskip

%+\addtocontents{toc}{{\it D. Reggiano}}
%+\label{ReggianoStart}

\begin{raggedright}  

Massimo Lenti\index{Lenti, M.}\\
INFN Sezione di Firenze, \\ 
Via G.Sansone 1, I-50019 Sesto F.(Firenze), ITALY
\bigskip\bigskip
\end{raggedright}

\section{Introduction}

Neutral Kaons were the place where CP violation was discovered for 
the first time~\cite{Christenson:1964fg}. Forty years after that discovery, kaons are still a privileged 
system where quarks coupling can be studied. The CKM quark mixing matrix~\cite{Kobayashi:1973fv} structure 
is usually addressed through the so-called unitarity triangle~\cite{Wolfenstein:1983yz}. The very popular picture 
of this triangle based on measurements coming from B mesons decays and mixing can be 
replaced by an analogous one based on Kaons decays. It is evident that the possibility 
to compare the theoretical prediction in two independent system (the B system and 
the K system) with comparable experimental and theoretical uncertainties is very promising 
in the view of any signal of new physics or of the possible failure of the present Standard Model of particle interactions.

The Kaon version of the unitarity triangle is mainly identified by the branching ratio of 
the $K^+\rightarrow\pi^+\nu\bar\nu$ decay and of the (direct CP-violating)
$K_L\rightarrow\pi^0\nu\bar\nu$, which are theoretically 
very clean even if very demanding from an experimental point of view.
The less theoretically clean $K_L\rightarrow\pi^0 e^+ e^-$ and $K_L\rightarrow\pi^0 \mu^+ \mu^-$ 
decays can be used in place of the $K_L\rightarrow\pi^0\nu\bar\nu$ decay,
but the different physical contributions to the decay must be disentangled.
The CP conserving part of $K_L\rightarrow\pi^0 l^+ l^-$ can be calculated from the branching ratio 
of the $K_L\rightarrow\pi^0\gamma\gamma$ decay measured by NA48~\cite{Lai:2002kf}:
\begin{equation}
 {\rm BR}(K_L\rightarrow\pi^0 e^+ e^-)_{\rm CP\,cons.} = 0.47^{+0.22}_{-0.18}\times 10^{-12}
\end{equation}
\begin{equation}
 {\rm BR}(K_L\rightarrow\pi^0 \mu^+ \mu^-)_{\rm CP\,cons.} \approx 10^{-12}
\end{equation}
The indirect CP violating part can be calculated using $K_S$ decays:
\begin{equation}
 {\rm BR}(K_L\rightarrow\pi^0 l^+ l^-)_{\rm ICP\,viol.} = |\epsilon|^2(\tau_L/\tau_S) {\rm BR}(K_S\rightarrow\pi^0 l^+ l^-)
\end{equation}
The  $K_S\rightarrow\pi^0 e^+ e^-$ and $K_S\rightarrow\pi^0 \mu^+ \mu^-$ decays are very important 
to constrain the direct CP violation part of ${\rm BR}(K_L\rightarrow\pi^0 l^+ l^-)$ which is expected to be 
of the order of few $10^{-12}$.

The basis of the unitarity triangle can be fixed by the $K_L\rightarrow\mu^+\mu^-$ decay, but many information 
are neeeded to disentangle the short-range and long-range contributions. 
The  $K_L\rightarrow e^+e^-\gamma$ and $K_L\rightarrow e^+e^- e^+e^-$ decays are very useful in this respect.

Hyperon decays are also very interesting for what concerns weak interactions. 
Hyperon decays can address parity violating modes like decay asymmetries, where some channels are very 
poorly known. The semileptonic hyperon decays can also measure the CKM parameter $|V_{us}|$ in an independent 
way from the most common method of using semileptonic Kaon decays.

\section{The beam setup}

%%%%%%%%%%%%%%%%%%%%%%%%%%%%%%%%%%%%%%%%%%%%%%%%%%%%%%%%%%%%%%%%%%%%%%%%%
%%
%%   use this format to include an .eps figure into your paper
%%
\begin{figure}[htb]
\begin{center}
\epsfig{file=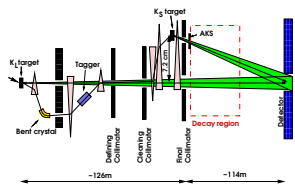,height=1.5in,angle=-90}
\caption{The NA48 beam setup.}
\label{fig:beams}
\end{center}
\end{figure}
%%%%%%%%%%%%%%%%%%%%%%%%%%%%%%%%%%%%%%%%%%%%%%%%%%%%%%%%%%%%%%%%%%%%%%%%%%%

NA48 is an experiment dedicated to the precise study 
of CP violation in the neutral Kaon system: its beam setup is made of 
two neutral beams, one dominated by $K_L$ decays, the other by $K_S$ 
decays. 
A primary proton beam from the CERN Super Proton Syncroton 
impinges, with a nominal flux of 1.5$\times 10^{12}$ particles per spill,
on a beryllium target (``Far Target'') 40 cm long and 2 mm wide; 
after a magnet sweeping sector and several stages of collimation, 
a neutral beam is formed. The exit face of the last collimator is 
located 126 m downstream of the target, at the entrance of the fiducial 
kaon decay region.

The primary protons which have not interacted in the target are deflected 
towards a bent silicon crystal: a small fraction of these protons is 
channeled by the crystal and form a secondary proton beam which is 
transported towards a second beryllium target (``Near Target''); after 
only 6 m of magnet sweeping and collimation another neutral beam 
is formed.

The two neutral beams can be present at the same time (``simultaneous beam 
runs'') or one per time (``Far Target run'' or ``Near Target run'') if 
only $K_L$ or $K_S$ decays have to be studied.

The secondary proton beam is usually at much lower intensity (about 
3$\times 10^7$ particles per spill) with respect to the primary one.
In some special runs (``High Intensity Near Target runs'') the primary proton 
beam is sent directly to the second target, removing the first target 
and by-passing the silicon crystal; the proton beam is attenuated and 
collimated to the desired intensity far upstream of the second target.

The neutral kaons beams can also be replaced by two opposite charged kaons beam.
\section{The Detectors}

%%%%%%%%%%%%%%%%%%%%%%%%%%%%%%%%%%%%%%%%%%%%%%%%%%%%%%%%%%%%%%%%%%%%%%%%%
%%
%%   use this format to include an .eps figure into your paper
%%
\begin{figure}[htb]
\begin{center}
\epsfig{file=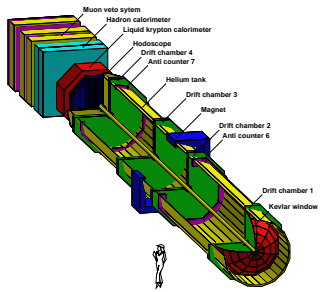,height=1.5in}
\caption{The NA48 detectors.}
\label{fig:detectors}
\end{center}
\end{figure}
%%%%%%%%%%%%%%%%%%%%%%%%%%%%%%%%%%%%%%%%%%%%%%%%%%%%%%%%%%%%%%%%%%%%%%%%%%%

The detectors are located downstream of the kaon decay volume which lies 
inside a large, 90 m long, vacuum tank terminated by a 0.3\% radiation 
lenghts thick Kevlar window; starting at the center of the Kevlar window, 
a 16 cm diameter vacuum beam pipe traverses all the detectors to let the 
neutral beam pass through vacuum.

Charged particles are detected using a high resolution magnetic 
spectrometer which consists of a dipole magnet with a horizontal 
transverse momentum kick of 265 MeV/c and a set of four drift chambers, 
two of them located upstream of the magnet and two downstream.
The magnetic spectrometer is contained inside a tank filled with helium 
in order to reduce multiple scattering. The momentum resolution is:
\begin{equation}
 \frac{\sigma_p}{p} (\%)= 0.48 \,\oplus\, 0.009\, p\;({\rm p\; in\; GeV/c})
\end{equation}  

A quasi-homogeneous liquid krypton (LKr) calorimeter is located 
downstream of the spectrometer. This detector has a 127 cm long projective 
tower structure which is made of copper-beryllium ribbons extending between
the front and the back of the calorimeter with a $\pm$48 mrad 
accordion geometry. The 13212 readout cells each have a cross-section 
of 2$\times$2 cm$^2$. The energy resolution is:
\begin{equation}
 \frac{\sigma_E}{E} (\%)= \frac{3.2}{\sqrt{E}} \,\oplus\, 
 \frac{9.0}{E} \,\oplus\, 0.42 \;({\rm E\; in\; GeV})
\end{equation}  

Two planes of scintillators, segmented in horizontal and vertical slabs,
form the charged hodoscope, located in between the magnetic spectrometer and the 
LKr calorimeter; it is used for triggering and measuring the time of 
charged particles.

Behind the LKr electromagnetic calorimeter, a 6.7 nuclear interaction lenght 
thick hadronic calorimeter is located, followed by a set of three planes 
of muon veto counters.

\begin{figure}[htb]
\begin{center}
\epsfig{file=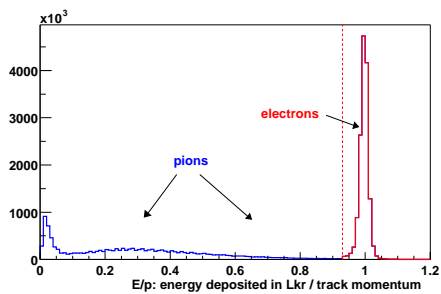,height=1.5in}
\caption{Ratio between the cluster energy measured by the LKr and the 
track momentum measured by the spectrometer.}
\label{fig:KLe3_eop}
\end{center}
\end{figure}

An electron (positron) is identified by a charged track, reconstructed by the magnetic 
spectrometer, whose extrapolation to the liquid krypton calorimeter matches a cluster in 
the LKr within 1.5 cm. The E/P ratio (ratio between the cluster energy measured by the LKr and the 
track momentum measured by the spectrometer) must be close to one; the usual cut is 
E/P$>$0.9 (see Figure~\ref{fig:KLe3_eop}). An electron-positron couple is selected requiring 
that their separation at the first drift chamber is larger than 2 cm in order to reject 
photon conversions.

%%%%%%%%%%%%%%%%%%%%%%%%%%%%%%%%%%%%%%%%%%%%%%%%%%%%%%%%%%%%%%%%%%%%%%%%%
%%
%%   use this format to include an .eps figure into your paper
%%
%%%%%%%%%%%%%%%%%%%%%%%%%%%%%%%%%%%%%%%%%%%%%%%%%%%%%%%%%%%%%%%%%%%%%%%%%%%

A proton or a charged pion is identified in a similar way as an electron but with 
an E/P$<$0.8 cut.
A muon is identified by a charged track whose extrapolation to the muon veto counters
matches at least one hit in the first two muon counters, with a coincidence gate of 4 ns.

A photon is identified by a a cluster in the liquid krypton calorimeter not matched 
by any charged track extrapolation.

A $\pi^0$ is identified by its main decay $\pi^0\rightarrow\gamma\gamma$, identifying 
two photons in the LKr. If the $\pi^0$ comes from a Kaon decay, the $\pi^0$ decay point is identified with the Kaon one,
reconstructed from the energy or momentum and position at the LKr of all the particles coming from the Kaon decay, 
imposing the Kaon mass. 

\section{Data samples}

Data were collected with the ``simultaneous beams'' 
setup in the years 1997, 1998, 1999 and 2001.

Data were collected in two days of 1999, in 40 days of 2000 and in all the 2002 run with the 
``High Intensity Near Target'' setup.  
A 40 days long ``Far Target run'' was also used in 2000.

In the years 2003 and 2004 charged kaons beams were used.

\section{$K_L\rightarrow e^+e^-\gamma$}

%%%%%%%%%%%%%%%%%%%%%%%%%%%%%%%%%%%%%%%%%%%%%%%%%%%%%%%%%%%%%%%%%%%%%%%%%
%%
%%   use this format to include an .eps figure into your paper
%%
\begin{figure}[htb]
\begin{center}
\epsfig{file=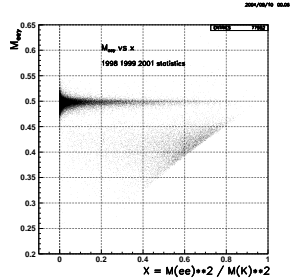,height=1.5in}
\caption{$e^+e^-\gamma$ invariant mass versus $x$; the $K_L\rightarrow e^+e^-\gamma$ signal is 
clearly visible.}
\label{fig:KLeeg_mee_meeg}
\end{center}
\end{figure}
%%%%%%%%%%%%%%%%%%%%%%%%%%%%%%%%%%%%%%%%%%%%%%%%%%%%%%%%%%%%%%%%%%%%%%%%%%%
The decay $K_L\rightarrow e^+e^-\gamma$ was identified by the detection 
of one electron-positron couple and a photon. 
The main background was given by 
$K_L\rightarrow \pi^0\pi^0_{\rm Dalitz}$ where the Dalitz decay is
$\pi^0_{\rm Dalitz}\rightarrow e^+e^-\gamma$; this background could mimic the 
signal if two photons were missed, but the invariant mass of the $e^+e^-\gamma$ was 
anyway smaller than the Kaon mass. Figure~\ref{fig:KLeeg_mee_meeg},
based on data collected in the year 1998, 1999 and 2001 shows the $e^+e^-\gamma$ invariant mass
versus the variable $x$, defined as the squared ratio of the $e^+ e^-$ invariant mass 
and the Kaon mass.

Using data collected only in the year 1997, corresponding to about 7000 signal events,
the branching ratio was measured~\cite{Fanti:1999rz}:
\begin{equation}
 {\rm BR}(K_L\rightarrow e^+e^-\gamma) = (1.06\pm0.02_{\rm stat}\pm0.02_{\rm syst}\pm0.04_{\rm norm})\times 10^{-5}
\end{equation}  
where the first error was statistical, the second systematic and the third one came from the uncertainty on the 
branching ratio of the normalization used ($K_L\rightarrow\pi^0\pi^0_{\rm Dalitz}$).
%%%%%%%%%%%%%%%%%%%%%%%%%%%%%%%%%%%%%%%%%%%%%%%%%%%%%%%%%%%%%%%%%%%%%%%%%
%%
%%   use this format to include an .eps figure into your paper
%%
\begin{figure}[htb]
\begin{center}
\epsfig{file=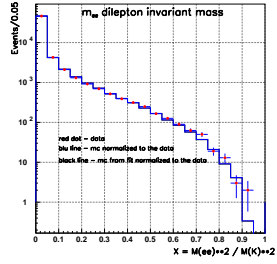,height=1.5in}
\caption{$x$ distribution for $K_L\rightarrow e^+e^-\gamma$ events; the form factor $\alpha_{K^*}$ fit is also shown.}
\label{fig:KLeeg_x}
\end{center}
\end{figure}
%%%%%%%%%%%%%%%%%%%%%%%%%%%%%%%%%%%%%%%%%%%%%%%%%%%%%%%%%%%%%%%%%%%%%%%%%%%
Using data collected in the year 1998, 1999 and 2001 with about 60000 candidates, it was possible to study the structure of the virtual 
photon producing the $e^+ e^-$ couple. In the BMS model~\cite{Bergstrom:1983rj} the form factor $\alpha_{K^*}$ measures the relative 
strength of the intermediate pseudoscalar and vector contribution. The form factor $\alpha_{K^*}$ can be fitted 
from the $x$ distribution of the signal events, as shown in Figure~\ref{fig:KLeeg_x}.

The result of the fit was:
\begin{equation}
 \alpha_{K^*} = -0.207\pm0.019_{\rm stat}\pm0.017_{\rm syst}[{\it preliminary}]
\end{equation}  

\section{$K_L\rightarrow e^+e^- e^+e^-$}

\begin{figure}[htb]
\begin{center}
\epsfig{file=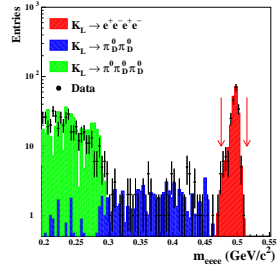,height=1.5in}
\caption{$e^+e^-e^+e^-$ invariant mass; the filled dots are the data, the histograms are the Monte Carlo prediction 
for the signal and the backgrounds. The signal region is shown by the arrows.}
\label{fig:KLeeee_mass}
\end{center}
\end{figure}

The $K_L\rightarrow e^+e^- e^+e^-$ decays is clearly related to the $K_L\rightarrow e^+e^-\gamma$ decay,
as also the second (virtual) photon gives rise to an $e^+e^-$ couple.
It was selected by identifying two electrons and two positrons and requiring that
the four tracks must come from a common decay vertex.
The main backgrounds were $K_L\rightarrow\pi^0\pi^0_{\rm Dalitz}\pi^0_{\rm Dalitz}$ and $K_L\rightarrow\pi^0_{\rm Dalitz}\pi^0_{\rm Dalitz}$,
but they were suppressed by a 475 MeV/c$^2<m_{e^+e^-e^+e^-}<$ 515 MeV/c$^2$ cut to a level smaller than $1\%$. 
Using data collected in the year 1998 and 1999, 200 signal events were selected, as can be seen 
in Figure~\ref{fig:KLeeee_mass}.
%%%%%%%%%%%%%%%%%%%%%%%%%%%%%%%%%%%%%%%%%%%%%%%%%%%%%%%%%%%%%%%%%%%%%%%%%
%%
%%   use this format to include an .eps figure into your paper
%%
%%%%%%%%%%%%%%%%%%%%%%%%%%%%%%%%%%%%%%%%%%%%%%%%%%%%%%%%%%%%%%%%%%%%%%%%%%%
The branching ratio was measured to be:
\begin{equation}
 {\rm BR}(K_L\rightarrow e^+e^-e^+e^-) = (3.30\pm0.24_{\rm stat}\pm0.14_{\rm syst}\pm0.10_{\rm norm})\times 10^{-8}[{\it preliminary}]
\end{equation}  
with the same meaning for the errors as in the previous section. The normalization channel used was 
$K_L\rightarrow\pi^+\pi^-\pi^0_{\rm Dalitz}$.

\section{$K_S\rightarrow \pi^0 e^+ e^-$}

The year 2002 data taking period was dedicated to rare $K_S$ decay with a ``High Intensity Near Target'' beam setup; in this data sample,
the $K_S\rightarrow \pi^0 e^+ e^-$ decay was looked for.
The $K_S\rightarrow\pi^0\pi^0$ decay with Dalitz decay of one of the $\pi^0$s or with a photon conversion or with 
a $\pi^0\rightarrow e^+e^-$ decay could mimic the signal. A cut with $m_{ee}>0.165$ GeV/c$^2$ rejected most of the backgrounds.
The $\pi^0$ was selected by a $|m_{\gamma\gamma}-M_{\pi^0}|<2.5$ MeV/c$^2$ cut. 
Finally a $|m_{e^+e^-\gamma\gamma}-M_{K}|<11.5$ MeV/c$^2$ cut was applied, where in this case the Kaon vertex was determined by the average position 
of closest distance of approach of the two charged tracks to the line joining the target and the Kaon center of gravity 
at the LKr.

$K_L$ were also present in the $K_S$ beam. The $K_L\rightarrow e^+e^-\gamma\gamma$ decay was a potentially dangerous background;
Using 2001 data (rich in $K_L$ decays) rescaled to 2002 data, a background of 0.075 $K_L\rightarrow e^+e^-\gamma\gamma$ events was 
estimated in the signal region.  

Another important possible source of backgrounds was given by the accidental overlap of fragments coming from two different decays.
Tracks and clusters were required to be in time within 3 ns and the accidental background was estimated using a control region 
where the time between fragment was in a range between 3 and 50 ns.

%%%%%%%%%%%%%%%%%%%%%%%%%%%%%%%%%%%%%%%%%%%%%%%%%%%%%%%%%%%%%%%%%%%%%%%%%
%%
%%   use this format to include an .eps figure into your paper
%%
\begin{figure}[htb]
\begin{center}
\epsfig{file=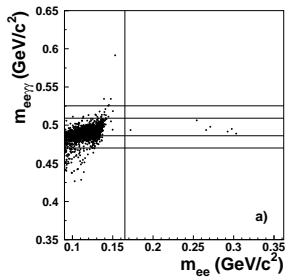,height=1.5in}
\caption{Scatter plot of $m_{e^+e^-\gamma\gamma}$ versus $m_{e^+e^-}$ for events passing all the cuts; 
the regions of 3$\sigma$ and 6$\sigma$ are shown}
\label{fig:mkaon_mee_intime_pm}
\end{center}
\end{figure}
%%%%%%%%%%%%%%%%%%%%%%%%%%%%%%%%%%%%%%%%%%%%%%%%%%%%%%%%%%%%%%%%%%%%%%%%%%%
%%%%%%%%%%%%%%%%%%%%%%%%%%%%%%%%%%%%%%%%%%%%%%%%%%%%%%%%%%%%%%%%%%%%%%%%%
%%
%%   use this format to include an .eps figure into your paper
%%
\begin{figure}[htb]
\begin{center}
\epsfig{file=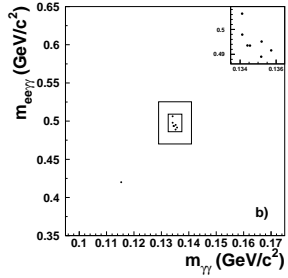,height=1.5in}
\caption{Scatter plot of $m_{e^+e^-\gamma\gamma}$ versus $m_{\gamma\gamma}$ for events passing all the cuts; 
the regions of 3$\sigma$ and 6$\sigma$ are shown}
\label{fig:mkaon_rmpi_intime_pm}
\end{center}
\end{figure}
%%%%%%%%%%%%%%%%%%%%%%%%%%%%%%%%%%%%%%%%%%%%%%%%%%%%%%%%%%%%%%%%%%%%%%%%%%%

Taking into account all the sources of backgrounds, $0.15^{+0.05}_{-0.04}$ events could mimic the signal.
Using a blind analysis approach, 7 $K_S\rightarrow \pi^0 e^+ e^-$ candidates were found, from which it was calculated~\cite{Batley:2003mu}:
\begin{equation}
 {\rm BR}(K_S\rightarrow \pi^0 e^+ e^-)_(m_{ee}>0.165\,{\rm GeV/c^2}) = (3.0^{+1.5}_{-1.2}(\rm stat) \pm 0.2(syst))\times 10^{-9}
\end{equation}
In order to extrapolate the result to any possible $m_{ee}$ value, a vector interaction with a unit form factor was assumed: 
\begin{equation}
 {\rm BR}(K_S\rightarrow \pi^0 e^+ e^-) = (5.8^{+2.8}_{-2.3}({\rm stat}) \pm 0.3({\rm syst}) \pm 0.8({\rm theor}))\times 10^{-9}
\end{equation}
where the last error came from the uncertainty on the theoretical model used for the extrapolation.

\section{$K_S\rightarrow \pi^0 \mu^+ \mu^-$}

%%%%%%%%%%%%%%%%%%%%%%%%%%%%%%%%%%%%%%%%%%%%%%%%%%%%%%%%%%%%%%%%%%%%%%%%%
%%
%%   use this format to include an .eps figure into your paper
%%
\begin{figure}[htb]
\begin{center}
\epsfig{file=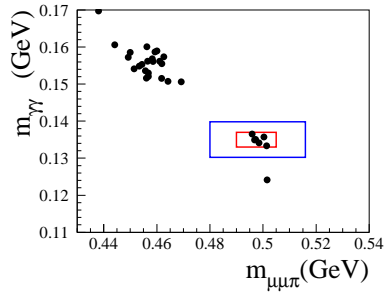,height=1.5in}
\caption{Scatter plot of $m_{\gamma\gamma}$ versus $m_{\mu^+\mu^-\pi^0}$ for events passing all the cuts; 
the regions of 2.5$\sigma$ and 6$\sigma$ are shown.}
\label{fig:paper_stplots_mkmpi0}
\end{center}
\end{figure}
%%%%%%%%%%%%%%%%%%%%%%%%%%%%%%%%%%%%%%%%%%%%%%%%%%%%%%%%%%%%%%%%%%%%%%%%%%%
%%%%%%%%%%%%%%%%%%%%%%%%%%%%%%%%%%%%%%%%%%%%%%%%%%%%%%%%%%%%%%%%%%%%%%%%%
%%
%%   use this format to include an .eps figure into your paper
%%
\begin{figure}[htb]
\begin{center}
\epsfig{file=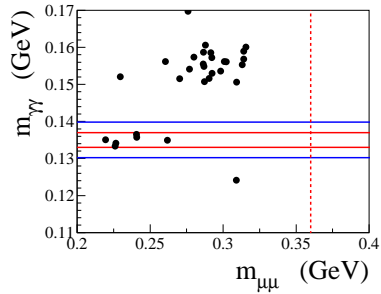,height=1.5in}
\caption{Scatter plot of $m_{\gamma\gamma}$ versus $m_{\mu^+\mu^-}$ for events passing all the cuts; 
the regions of 2.5$\sigma$ and 6$\sigma$ are shown; the $m_{\mu^+\mu^-}$ kinematic limit is also shown.}
\label{fig:paper_stplots_mpi0_intime_nomkcut}
\end{center}
\end{figure}
%%%%%%%%%%%%%%%%%%%%%%%%%%%%%%%%%%%%%%%%%%%%%%%%%%%%%%%%%%%%%%%%%%%%%%%%%%%
The $K_S\rightarrow \pi^0 \mu^+ \mu^-$ is the same as the decay discussed in the previous section but with the $e^+e^-$ 
replaced by two opposite charged muons. 
The $\pi^0$ was selected by a $|m_{\gamma\gamma}-M_{\pi^0}|<2.0$ MeV/c$^2$ cut and  
a $|m_{\mu^+\mu^-\gamma\gamma}-M_{K}|<7.5$ MeV/c$^2$ cut was applied (see the previous section for the definition of vertex position).

The main physical background was given by $K_L\rightarrow\pi^+\pi^-\pi^0$ with pion decay in flight which was anyway suppressed 
by the total invariant mass cut.
The $K_L\rightarrow \mu^+\mu^-\gamma\gamma$ decay was suppressed by its low branching ratio (of the order of $10^{-9}$) and 
by the pion mass cut.

The background from accidental overlap of different decay was checked in a control region with time difference between 3 and 60 ns 
between fragments, extrapolated to the signal region of $\pm 1.5$ ns.

6 events were found with an expected background of $0.22^{+0.19}_{-0.12}$ events in the 2002 data sample, corresponding to~\cite{Batley:2004wg}:  
\begin{equation}
 {\rm BR}(K_S\rightarrow \pi^0 \mu^+ \mu^-) = (2.9^{+1.5}_{-1.2}({\rm stat}) \pm 0.2({\rm syst})\times 10^{-9}
\end{equation}

\section{Comments on $K_S\rightarrow\pi^0 l^+l^-$}
Chiral perturbation theory can be used to predict the BR($K_S\rightarrow\pi^0 l^+l^-$) as a function of two parameters,
$a_S$ and $b_S$~\cite{D'Ambrosio:1998yj}:
\begin{equation}
 {\rm BR}(K_S\rightarrow\pi^0 e^+e^-)     = [0.01 - 0.76a_S - 0.21b_S + 46.5a_S^2 + 12.9a_Sb_S + 1.44b_S^2] \times 10^{-10}
\end{equation}
\begin{equation}
 {\rm BR}(K_S\rightarrow\pi^0 \mu^+\mu^-) = [0.07 - 4.52a_S - 1.50b_S + 98.7a_S^2 + 57.7a_Sb_S + 8.95b_S^2] \times 10^{-11}
\end{equation}
Within the VMD model~\cite{D'Ambrosio:1998yj}, which predicts $b_S=0.4 a_S$ the value of $|a_S|$ can be derived: 
$|a_S|=1.06^{+0.26}_{-0.21}\pm0.07$ from ${\rm BR}(K_S\rightarrow\pi^0 e^+e^-)$ and 
$|a_S|=1.54^{+0.40}_{-0.32}\pm0.06$ from ${\rm BR}(K_S\rightarrow\pi^0 \mu^+\mu^-)$.

The central value for the total branching ratio of $K_L\rightarrow\pi^0 e^+e^-$ (CP conserving part, 
indirect CP violating, direct CP violating and interference term) was estimated to be $32\times 10^{-12}$ or 
$12\times 10^{-12}$, depending on the sign of the interference term. The central value of 
${\rm BR}(K_L\rightarrow\pi^0 \mu^+\mu^-)$ was estimated to be $19\times 10^{-12}$ or 
$13\times 10^{-12}$ for the same sign ambiguity.

\section{$\Xi^0\rightarrow\Lambda\gamma$}

\begin{figure}[htb]
\begin{center}
\epsfig{file=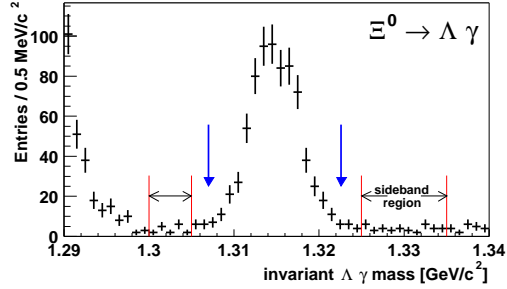,height=1.5in}
\caption{$\Lambda\gamma$ invariant mass; the signal and sidebands regions are indicated}
\label{fig:XiMass_descriptions}
\end{center}
\end{figure}

Using the 1999 ``High Intensity Near Target'' run, 730 $\Xi^0\rightarrow\Lambda\gamma$ candidates 
were selected with $\Lambda\rightarrow p\pi^-$.
The selection was based on two opposite charged tracks with an invariant mass within 2.7 MeV/c$^2$ from the 
$\Lambda$ mass and a cluster in the calorimeter with at least 15 GeV energy, separated by at least 25 cm from 
each track extrapolated point.
The expected background was estimated from the sidebands in the $\Lambda\gamma$ invariant mass 
(see Figure~\ref{fig:XiMass_descriptions}) to be $58.2\pm7.8$ events and~\cite{Lai:2004ak}
\begin{equation}
 {\rm BR}(\Xi^0\rightarrow\Lambda\gamma) = (1.16\pm0.05_{\rm stat}\pm0.06_{\rm syst})\times 10^{-3}
\end{equation}
%%%%%%%%%%%%%%%%%%%%%%%%%%%%%%%%%%%%%%%%%%%%%%%%%%%%%%%%%%%%%%%%%%%%%%%%%
%%
%%   use this format to include an .eps figure into your paper
%%
%%%%%%%%%%%%%%%%%%%%%%%%%%%%%%%%%%%%%%%%%%%%%%%%%%%%%%%%%%%%%%%%%%%%%%%%%%%

\begin{figure}[htb]
\begin{center}
\epsfig{file=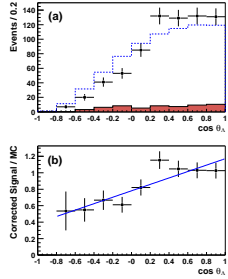,height=1.5in}
\caption{$\Xi^0\rightarrow\Lambda\gamma$ decay asymmetry fit to the $\Theta_{\Lambda}$ angle distribution}
\label{fig:AsymmetryFit}
\end{center}
\end{figure}

The $\Xi^0\rightarrow\Lambda\gamma$ decay asymmetry $\alpha(\Xi^0\rightarrow\Lambda\gamma)$ can be fitted from the distribution of the 
$\Theta_{\Lambda}$ angle, defined in the $\Lambda$ rest frame as the angle between the 
incoming $\Xi^0$ and the outgoing proton as can be seen in Figure~\ref{fig:AsymmetryFit}.
%%%%%%%%%%%%%%%%%%%%%%%%%%%%%%%%%%%%%%%%%%%%%%%%%%%%%%%%%%%%%%%%%%%%%%%%%
%%
%%   use this format to include an .eps figure into your paper
%%
%%%%%%%%%%%%%%%%%%%%%%%%%%%%%%%%%%%%%%%%%%%%%%%%%%%%%%%%%%%%%%%%%%%%%%%%%%%
The result of the fit was~\cite{Lai:2004ak}:
\begin{equation}
 \alpha(\Xi^0\rightarrow\Lambda\gamma) = -0.78\pm0.18_{\rm stat}\pm0.06_{\rm syst}
\end{equation}

\section{$\Xi^0\rightarrow\Sigma^+ e^-\bar\nu$}

\begin{figure}[htb]
\begin{center}
\epsfig{file=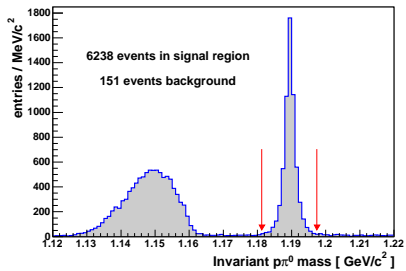,height=1.5in}
\caption{$p\pi^0$ invariant mass for $\Xi^0\rightarrow\Sigma^+ e^-\bar\nu$ candidates; the arrows indicate the signal region}
\label{fig:Sigma_mass}
\end{center}
\end{figure}

The $\Xi^0\rightarrow\Sigma^+ e^-\bar\nu$ decay was looked for in the 2002 ``High Intensity Near Target'' run.
The $\Sigma^+$ was reconstructed by its $p\pi^0$ decay. 
The $\Xi^0$ beta decay was the only source of $\Sigma^+$ in the neutral beam of the 2002 NA48 beam setup.
If an electron was found together with the $\Sigma^+$, the $\Xi^0$ beta decay was identified, as can be seen from Figure~\ref{fig:Sigma_mass}.
%%%%%%%%%%%%%%%%%%%%%%%%%%%%%%%%%%%%%%%%%%%%%%%%%%%%%%%%%%%%%%%%%%%%%%%%%
%%
%%   use this format to include an .eps figure into your paper
%%
%%%%%%%%%%%%%%%%%%%%%%%%%%%%%%%%%%%%%%%%%%%%%%%%%%%%%%%%%%%%%%%%%%%%%%%%%%%
6238 signal candidates were found with a background of about 150 events estimated from the mass sidebands,
from which the branching ratio was measured:
\begin{equation}
 {\rm BR}(\Xi^0\rightarrow\Sigma^+ e^-\bar\nu) = (2.51\pm0.03_{\rm stat}\pm0.11_{\rm syst})\times 10^{-4}[{\it preliminary}]
\end{equation}
In order to derive the CKM element $V_{us}$ some assumption on the form factors must be used:
we chose $f_1(0)=1$ and $g_1/f_1 = 1.32^{+0.21}_{0.17}\pm 0.05$~\cite{Alavi-Harati:2001xk} 
from which we calculated:
\begin{equation}
 |V_{us}| = 0.214\pm0.06^{+0.030}_{-0.025}[{\it preliminary}]
\end{equation}

\section{Conclusions}
The NA48 experiment provided a lot of measurement concerning rare Kaon and Hyperon decays which allowed important 
constrains of the CKM paradigm in an indipendent way from the B meson system. Many other channels are presently 
under investigation by the NA48 collaboration and their measurement is foreseen in the near future.

\bigskip

It is a pleasure to thanks the organizers and the speakers of the HEP-MAD04 conference held in Antananarivo 
from 27th September to 3rd October 2004 for the friendly atmosphere and the fruitful discussions.

\def\Discussion{
\setlength{\parskip}{0.3cm}\setlength{\parindent}{0.0cm}
     \bigskip\bigskip      {\Large {\bf Discussion}} \bigskip}
\def\speaker#1{{\bf #1:}\ }
\def\endDiscussion{}

\end{document}

%% file: econfmacros.tex
\def\beq{\begin{equation}}
\def\eeq#1{\label{#1}\end{equation}}
\def\eeqn{\end{equation}}

\def\beqa{\begin{eqnarray}}
\def\eeqa#1{\label{#1}\end{eqnarray}}
\def\eeqan{\end{eqnarray}}

\let\bar=\overbar

\def\Dslash{\not{\hbox{\kern-4pt $D$}}}
\def\dslash{\not{\hbox{\kern-2pt $\del$}}}

\def\msb{{\bar{\ssstyle M \kern -1pt S}}}